\documentclass{iaus}
\usepackage{graphicx}

\title[Chemical spots on HgMn stars] 
{Chemical spots and their dynamical evolution on HgMn stars}

\author[Korhonen et al.]   
{Heidi Korhonen$^1$, Swetlana Hubrig$^2$, Maryline Briquet$^3$, Federico 
  Gonz{\'a}lez$^4$, Igor Savanov$^{5}$}

\affiliation{$^1$ European Southern Observatory, Karl-Schwarzschild-Str 2, 
  D-85748, Garching bei M\"unchen, Germany\\
  email: {\tt hkorhone@eso.org}\\[\affilskip]
  $^2$ Astrophysikalisches Institut Potsdam, An der Sternwarte 16, D-14482 
  Potsdam, Germany\\[\affilskip]
  $^3$ Instituut voor Sterrenkunde, Katholieke Universiteit Leuven, 
  Celestijnenlaan 200 D, B-3001 Leuven, Belgium\\[\affilskip]
  $^4$ Instituto de Ciencias Astronomicas, de la Tierra, y del Espacio (ICATE),
  5400 San Juan, Argentina\\[\affilskip]
  $^5$ Institute of Astronomy, Russian Academy of Sciences, Pyatnitskaya 48, 
  Moscow 119017, Russia	
}

\pubyear{2010}
\volume{273}  
\pagerange{119--126}
\jname{Title of your IAU Symposium}
\editors{A.C. Editor, B.D. Editor \& C.E. Editor, eds.}
\begin{document}

\maketitle

\begin{abstract}
Our recent studies of late B-type stars with HgMn peculiarity revealed for the 
first time the presence of fast dynamical evolution of chemical spots on their 
surfaces. These observations suggest a hitherto unknown physical process 
operating in the stars with radiative outer envelopes. Furthermore, we have 
also discovered existence of magnetic fields on these stars that have up to now
been thought to be non-magnetic. Here we will discuss the dynamical spot 
evolution on HD~11753 and our new results on magnetic fields on AR~Aur.
\end{abstract}

\firstsection 

\section{Introduction}

Recently, \cite[Kochukhov et al.\,(2007)]{koch07}, using Doppler Imaging
technique, reported a discovery of secular evolution of the mercury
distribution on the surface of the HgMn star $\alpha$~And. However, this
result was never verified by other studies due to the lack of observational
data. Until very recently, the only other HgMn star with a published surface
elemental distribution was AR\,Aur (\cite[Hubrig et al.\,2006]{hub06}), 
where the discovered surface chemical inhomogeneities are related to the 
relative position of the companion star. The elements Y and Sr are strongly 
concentrated in an equatorial ring, which has a gap exactly on the area 
permanently facing the secondary.

A large number of spectra of a sample of HgMn stars were obtained with the
CORALIE spectrograph at the 1.2~m Euler telescope on La Silla during a
programme dedicated to a search for SPB-like pulsations in B-type stars. In 
\cite[Briquet et al. (2010)]{briq} we published two sets of surface maps of 
HD\,11753 based on these data. The maps are separated by approximately two 
months and are obtained from three different elements: \mbox{Y\,{\sc ii}}, 
\mbox{Ti\,{\sc ii}} and \mbox{Sr\,{\sc ii}}. The maps made from the 
\mbox{Y\,{\sc ii}} 4900~{\AA} line (see Fig.~\ref{fig1}) exhibit a high 
abundance region at phases 0.5--1.0 extending from the latitude 45$^{\circ}$ to
the pole, with an extension to the equator around the phase 0.8. The 
\mbox{Y\,{\sc ii}} abundance distribution shows also a high latitude lower 
abundance spot around phases 0.2--0.4. Clear evolution in the surface features 
is present during the two months that separate the datasets. The lower 
abundance high latitude feature at phases 0.2--0.4 becomes more extended and 
less prominent in the second set, while the abundance of the high abundance 
spot at phases 0.6--1.0 gets more prominent with time.

We have also obtained measurements of the magnetic field strength with the 
moment technique using several elements in a circularly polarised high 
resolution spectrum of another HgMn star, AR~Aur. These observations revealed 
the presence of a longitudinal magnetic field in both stellar components 
(\cite[Hubrig et al.\,2010]{hub10}). 

Here, we continue the investigation of HD~11753 using newer CORALIE data from 
2009 and 2010, and discuss the magnetic field measurements of AR~Aur.

\section{Spot evolution in HD11753 for 2000--2010}

HD~11753 is a single-lined spectroscopic binary with an effective temperature 
of 10612~K (\cite[Dolk et al.\,2003]{dolk2003}). According to our observations 
the orbital period of the binary would be long, and the projected rotational 
velocity v$\sin i$=13.5km/s (\cite[Briquet et al. 2010]{briq}). After adding 
the new 2009 and 2010 datasets the rotational period is improved to P=9.531\,d 
(\cite[Korhonen et al. 2010]{kor10}).

We have obtained Doppler images of HD~11753 from CORALIE spectra for four 
different epochs. In Doppler imaging spectroscopic observations at different 
rotational phases are used to measure the rotationally modulated distortions in
the line-profiles. These distortions are produced by the inhomogeneous 
distribution of a surface characteristic, e.g., surface temperature or element 
abundance. Surface maps are constructed by combining all the observations from 
different phases and comparing them with synthetic model line-profiles. For 
accurate Doppler imaging the shape and changes of the line-profile have to be 
well defined. This requires high resolving power and high 
signal-to-noise-ratio.
 
\begin{figure}
\begin{center}
 \includegraphics[width=2.6in]{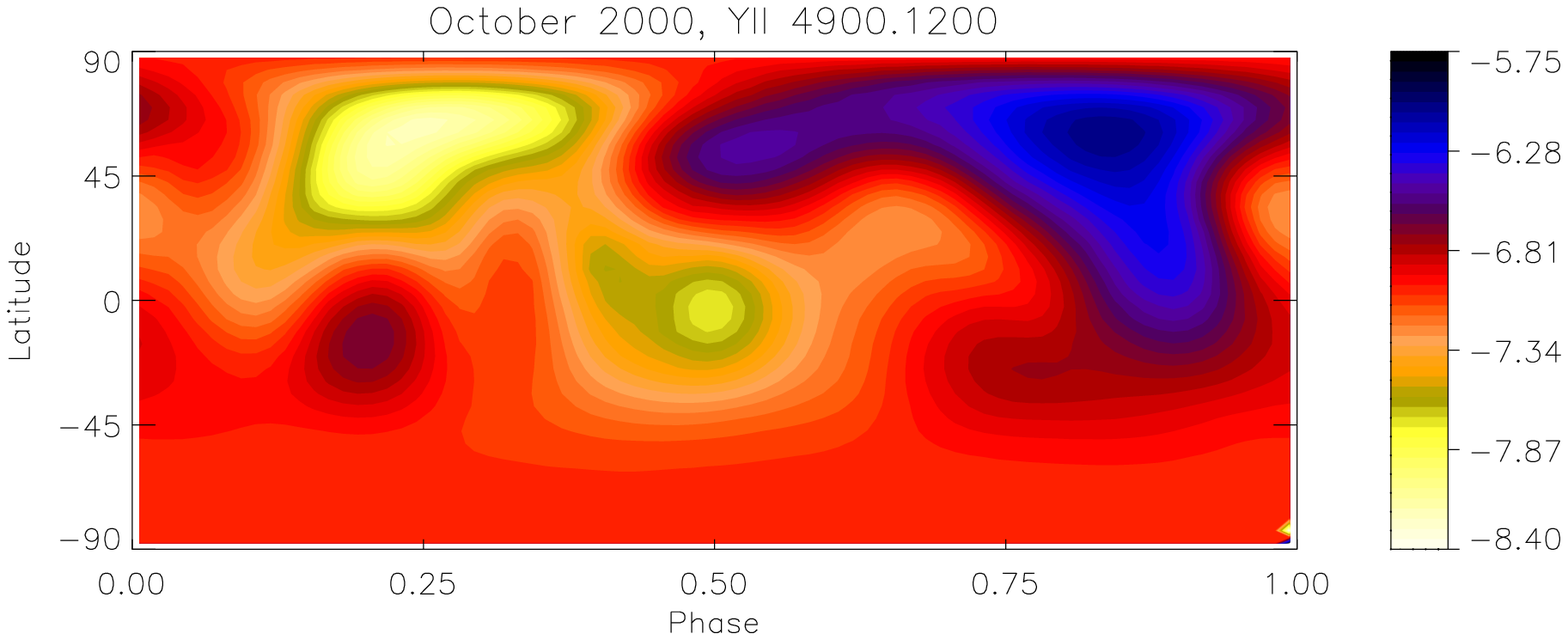}
 \includegraphics[width=2.6in]{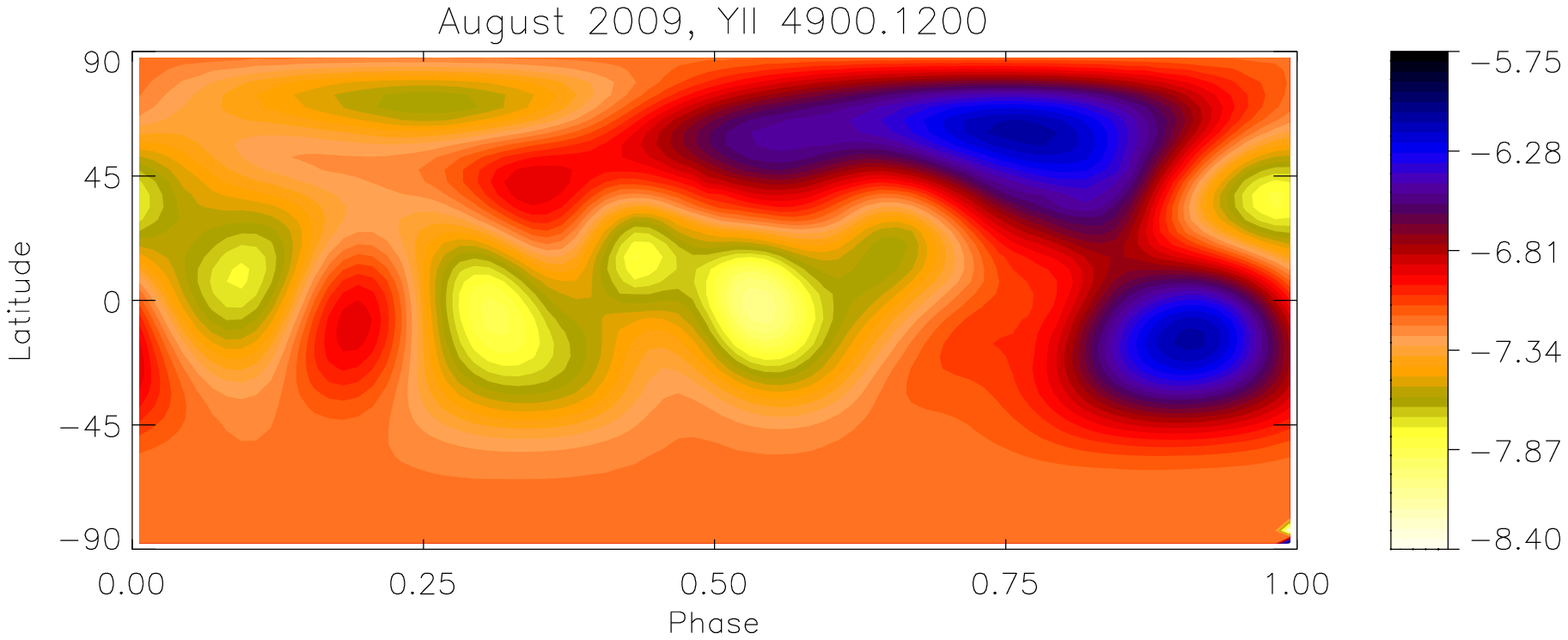}
 \includegraphics[width=2.6in]{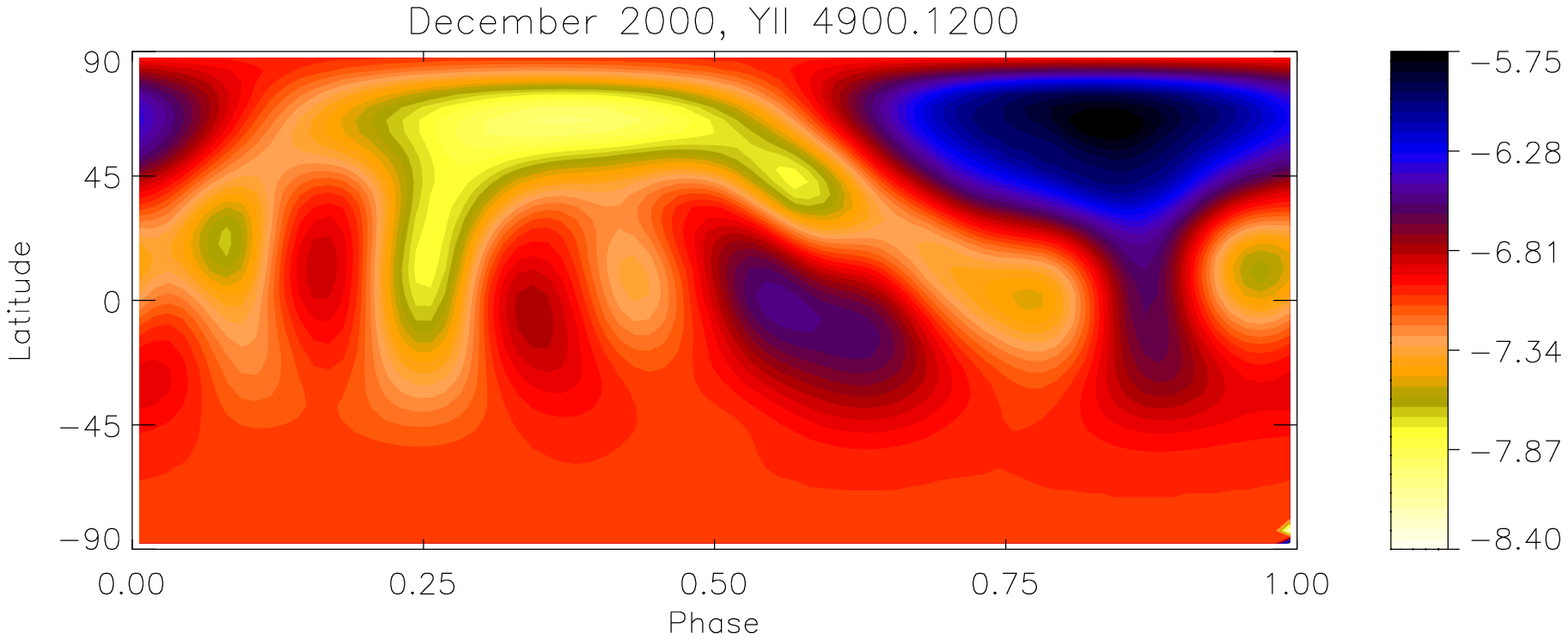}
 \includegraphics[width=2.6in]{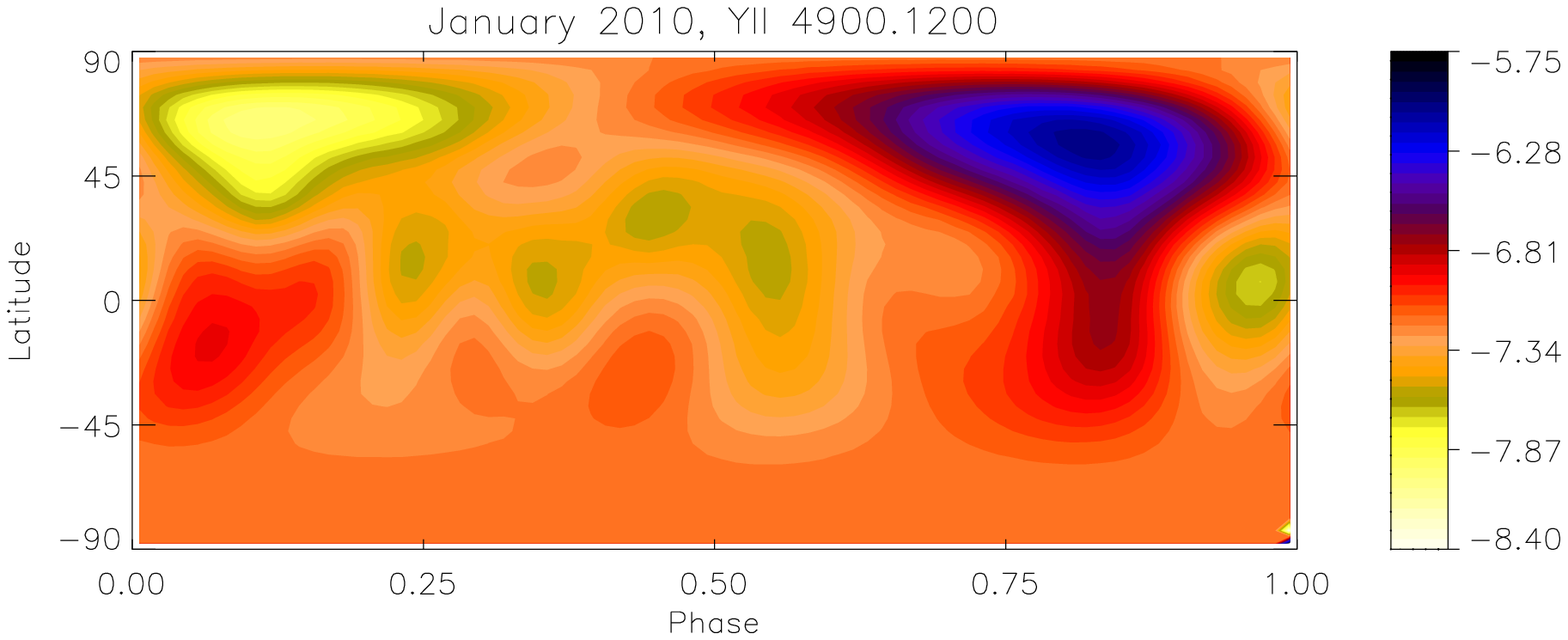}
 \caption{Chemical surface maps of HD~11753 from the \mbox{Y\,{\sc ii}} 
     4900~{\AA} line at four different epochs. In the maps the abscissa is the 
     longitude in phases and the ordinate is the latitude in degrees. Colour 
     indicates different abundances, with darker denoting higher abundance.}
   \label{fig1}
\end{center}
\end{figure}

The Doppler images of HD~11753 using the \mbox{Y\,{\sc ii}} 4900~{\AA} line are
shown in Fig.\,\ref{fig1}. The two first maps, both from 2000, have 65 days in 
between them. Clear temporal evolution of the chemical spots occurs even on 
such short timescales. The high abundance spot around the phase 0.75 gets more 
concentrated with time, and the lower abundance spot around the phase 0.25 
more extended. These two maps were already published by \cite[Briquet et al. 
(2010)]{briq}, but the August 2009 and January 2010 maps are previously 
unpublished. These latter maps have approximately four and half months in 
between them, and again they show temporal evolution of the surface structures.
The high abundance spot of phase 0.75 is at high latitudes much more extended 
in August 2009 than in January 2010. Also, the equatorial high abundance spot 
seen at phase 0.85 in the August 2009 map has disappeared before January 2010. 
The lower abundance spot of phase 0.0--0.4 is almost non-existent in August 
2009, but clearly present in January 2010. However, the August 2009 dataset has
a large phase gap, 0.17--0.47, close to the phase of the lower abundance spot. 
Our tests show, though, that a phase gap of 0.3 in phase does not affect the 
recovery of such large surface features (see 
\cite[Korhonen et al. 2010]{kor10}).

All in all, the chemical spots retain their position on the stellar surface 
stably the almost 10 year period our observations cover. The exact shape 
changes, though, and this change happens even on time scales of months.

\section{Magnetic field in AR~Aur}

The double-lined spectroscopic binary AR\,Aur has an orbital period of 4.13\,d.
It is a young system with an age of only 4$\times{}$10$^6$\,yr and its primary,
showing HgMn peculiarity, is exactly on the Zero Age Main Sequence while the 
secondary is still contracting towards it. Variability of spectral lines 
associated with a large number of chemical elements was reported for the first 
time for the primary component of this eclipsing binary by 
\cite[Hubrig et al.\,(2006)]{hub06}. 

Doppler maps for the elements Mn, Sr, Y, and Hg using nine spectra of AR\,Aur 
observed at the European Southern Observatory with the UVES spectrograph at 
UT2 in 2005 were for the first time presented at the IAU Symposium~259 by 
\cite[Savanov et al. (2009)]{sav09}. To prove the presence of a dynamical 
evolution of spots also on the surface of AR\,Aur, we obtained new 
spectroscopic data with the Coud\'e Spectrograph of the 2.0\,m telescope of the
Th\"uringer Landessternwarte and the SES spectrograph of the  1.2\,m STELLA-I 
robotic telescope at the Teide Observatory. A number of SOFIN spectra of
AR Aur were obtained in 2002 at the Nordic Optical Telescope, which we also 
used in our analysis. Our new results show secular evolution of the chemical 
spots on AR~Aur (\cite[Hubrig et al.\,2010]{hub10}). Fig.~\ref{fig2} shows 
{Sr\,{\sc ii} 4215.5~{\AA} line during three different epochs (late 2002, late 
2005, and late 2008 -- early 2009), but at the same rotational phase, 
$\sim$0.8. The shapes of the lineprofiles are clearly different indicating that
also the Sr spots very likely changed their shape and abundance with time.

\begin{figure}
\begin{center}
 \includegraphics[width=2.5in]{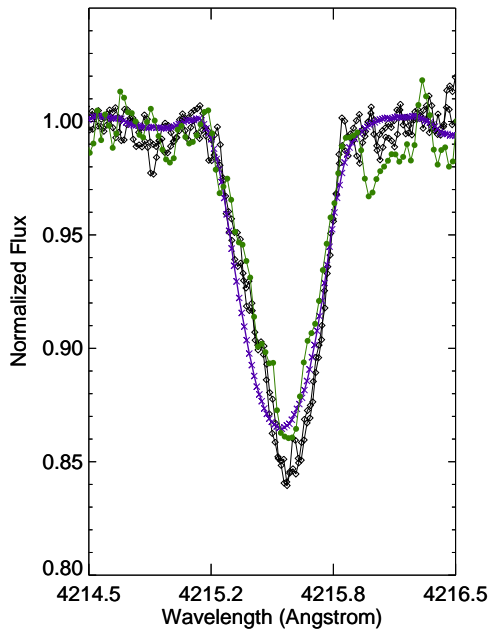}
 \caption{Line profiles of the Sr\,{\sc ii} 4215.5 {\AA} line at the rotation 
   phase $\sim$0.8 for late 2002 (open diamonds), late 2005 (crosses), and late
   2009 -- early 2010 (filled circles) show clearly different shapes.}
   \label{fig2}
\end{center}
\end{figure}

To pinpoint the mechanism responsible for the surface structure formation in 
HgMn stars, we carried out spectropolarimetric observations of AR\,Aur and 
investigated the presence of a magnetic field during a rotational phase of very
good visibility of the spots of overabundant elements. The spectropolarimetric 
observations of AR\,Aur at the rotation phase 0.622 were obtained with the 
low-resolution camera of SOFIN ($R\approx$30\,000) at the Nordic Optical 
Telescope. Since most elements are expected to be inhomogeneously distributed
over the surface of the primary of AR\,Aur, magnetic field measurements were 
carried out for samples of Ti, Cr, Fe, and Y lines separately. Among the 
elements showing line variability, the selected elements have numerous 
transitions in the observed optical spectral region, allowing us to sort out 
the best samples of clean unblended spectral lines with different Land{\'e} 
factors.

Our magnetic field measurements, which are discussed in detail by \cite[Hubrig 
et al.\,(2010)]{hub10}, were done using the formalism described by 
\cite[Mathys (1994)]{Mathys94}. A longitudinal magnetic field at a level higher
than 3$\sigma$ of the order of a few hundred Gauss is detected in 
\mbox{Fe\,{\sc ii}}, \mbox{Ti\,{\sc ii}}, and \mbox{Y\,{\sc ii}} lines, while a
quadratic magnetic field $\left<B\right>$=$8284\pm1501$\,G at 5.5$\sigma$ level
was measured in \mbox{Ti\,{\sc ii}} lines. No crossover at 3$\sigma$ confidence
level was detected for the elements studied. Further, we detect a weak 
longitudinal magnetic field, $\left<B_z\right>$=$-$$229\pm56$\,G, in the 
secondary component using a sample of nine \mbox{Fe\,{\sc ii}} lines. The main 
limitation on the accuracy achieved in our determinations is set by the small
number of lines that can be used for magnetic field measurements. The diagnosis
of the quadratic field is more difficult than that of the longitudinal magnetic
field, and it depends much more critically on the number of lines that can be 
used for the analysis. 

\section{Summary}

The fast dynamic evolution of the spots on HD~11753 implies hitherto unknown 
physical mechanism operating in the outer envelopes of late B-type stars with 
HgMn peculiarity and the detection of the magnetic field in AR~Aur shatters 
the traditional view that HgMn stars do not exhibit magnetic fields. For the
proper understanding of the nature of these stars we need accurate information
on the element spot configuration and underlying magnetic fields in a sample
of HgMn stars to determine a link between these properties and stellar
fundamental parameters such as rotation rate, temperature, evolutionary state,
stellar mass, multiplicity and orbital parameters. It is clear that time series
of high resolution spectropolarimetric observations are needed to solve the 
puzzle these stars represent.

\end{document}